\documentclass[11pt]{article}

\usepackage[final]{acl}

\usepackage{times}
\usepackage{latexsym}

\usepackage[T1]{fontenc}
\usepackage[utf8]{inputenc}

\usepackage{microtype}

\usepackage{inconsolata}

\usepackage{graphicx}

\usepackage{amsmath}
\usepackage{booktabs}
\usepackage{xcolor}
\usepackage{float}
\title{BSCodec: A Band-Split Neural Codec for High-Quality Universal Audio Reconstruction}

\author{
 \textbf{Haoran Wang\textsuperscript{1,2}},
 \textbf{Jiatong Shi\textsuperscript{1}},
 \textbf{Jinchuan Tian\textsuperscript{1}},
 \textbf{Bohan Li\textsuperscript{2}},
\\
 \textbf{Kai Yu\textsuperscript{2}},
 \textbf{Shinji Watanabe\textsuperscript{1}\thanks{Corresponding author.}}
\\
 \textsuperscript{1}Carnegie Mellon University~~
 \textsuperscript{2}Shanghai Jiao Tong University
\\
}

\begin{document}
\maketitle
\begin{abstract}

Neural audio codecs have recently enabled high-fidelity reconstruction at high compression rates, especially for speech. However, speech and non-speech audio exhibit fundamentally different spectral characteristics: speech energy concentrates in narrow bands around pitch harmonics (80-400 Hz), while non-speech audio requires faithful reproduction across the full spectrum, particularly preserving higher frequencies that define timbre and texture. This poses a challenge—speech-optimized neural codecs suffer degradation on music or sound. Treating the full spectrum holistically is suboptimal: frequency bands have vastly different information density and perceptual importance by content type, yet full-band approaches apply uniform capacity across frequencies without accounting for these acoustic structures. To address this gap, we propose \textbf{BSCodec} (Band-Split Codec), a novel neural audio codec architecture that splits the spectral dimension into separate bands and compresses each band independently. Experimental results demonstrate that BSCodec achieves superior reconstruction over baselines across sound and music, while maintaining competitive quality in the speech domain, when trained on the same combined dataset of speech, music and sound. Downstream benchmark tasks further confirm that BSCodec shows strong potential for use in downstream applications.\footnote{https://github.com/whr-a/espnet/tree/bscodec}
\end{abstract}
% https://anonymous.4open.science/r/espnet-2C03

% We observe that uniform compression across all frequencies is suboptimal: different frequency bands have vastly different information density and perceptual importance depending on content type.

\section{Introduction}
\label{sec:inc}

Neural audio codecs (NACs)~\cite{mousavi2025discrete, guo2025recent, shi2024espnet, defossez2022high, kumar2023high, zeghidour2021soundstream} have revolutionized audio compression and achieved high-fidelity reconstruction. These codecs generate discrete tokens that not only enable efficient transmission but also serve as representations for downstream audio understanding tasks~\cite{wang2023neural, tian2025espnet}. To enhance task-specific performance, recent work~\cite{zhang2023speechtokenizer, ye2025codec} has incorporated external semantic information into codec design, demonstrating improved results on speech-related benchmarks. Additionally, single-codebook approaches~\cite{ji2024wavtokenizer, xin2024bigcodec, jiang2025unicodec} have achieved excellent reconstruction quality and downstream performance on clean speech through carefully designed codebook structures, encoder-decoder architectures, and other components.

\begin{figure}[t]
\centering
\includegraphics[width=\columnwidth]{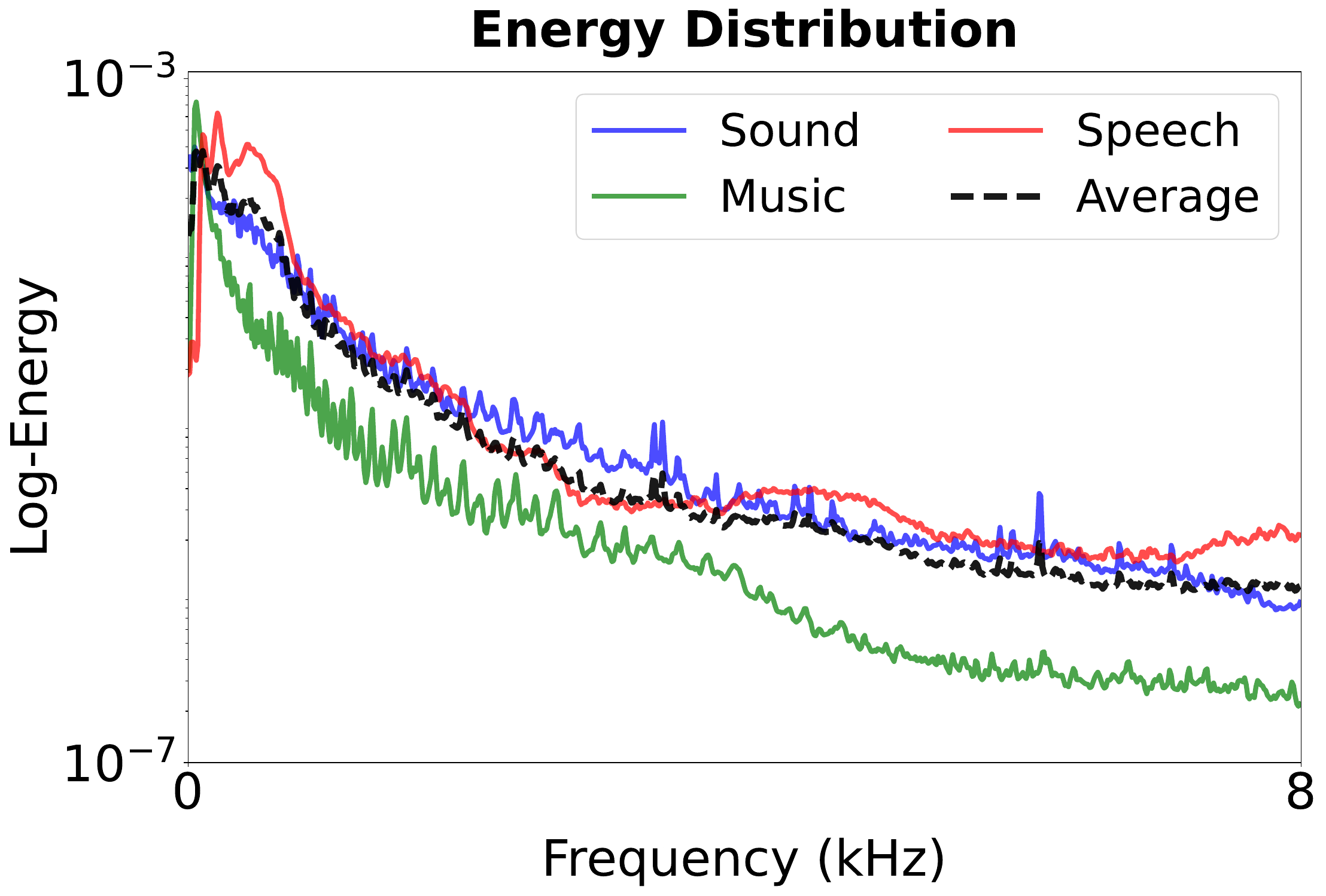}
\caption{Frequency domain energy distribution of speech, sound and music (extraction method in Appendix~\ref{app:energy_extraction}). The distributions exhibit significant structural differences across these three domains, demonstrating distinct spectral characteristics.}
\label{fig:energy_spectrum}
\end{figure}

However, the real acoustic world would contain components beyond speech, such as sound and music, all with vastly different acoustic characteristics. As illustrated in Figure~\ref{fig:energy_spectrum}, speech, music, and sound exhibit substantially different average energy distributions across the frequency spectrum. While the aforementioned task-oriented codecs excel on clean speech, their performance degrades significantly~\cite{mousavi2025discrete} when applied to music and sound domains. Given these spectral differences shown in Figure~\ref{fig:energy_spectrum}, it is reasonable that speech codec designs struggle to generalize to music and sound. Current RVQ-based generalized codecs like DAC~\cite{kumar2023high} still exhibit advantages~\cite{mousavi2025discrete} in reconstruction quality for universal audio.

While RVQ-based codecs demonstrate reasonable multi-domain performance, their design could be enhanced. The residual quantization hierarchy lacks explicit acoustic structure, with each layer quantizing residuals that do not correspond to interpretable acoustic attributes. This poses challenges for multi-domain compression: as different audio types have distinct energy distributions (Figure~\ref{fig:energy_spectrum}), a codec handling multiple domains faces a signal source with substantially higher entropy, demanding increased bitrate. Also RVQ applies uniform residual encoding across all domains, requiring the encoder to learn entangled representations for diverse content. Alternative quantization schemes like FSQ~\cite{mentzer2023finite} and GroupVQ~\cite{yang2023hifi} partition along dimensional axes, but these divisions similarly lack grounding in the physical properties of audio signals.

To effectively handle diverse audio domains, we need to account for their structural differences. As shown in Figure~\ref{fig:energy_spectrum}, different domains exhibit substantially different energy distribution across frequency bands. Therefore, band-splitting offers a natural way to decouple these domain-specific structures by processing each frequency band independently. This approach has proven effective in other audio tasks~\cite{luo2023music, yang2021multi}, and a RVQ-based multi-band codec has been successfully built for the speech domain~\cite{ng2025multiband}, which motivate us to adopt similar decomposition principles for universal audio codec design.

Motivated by these insights, we propose a band-split codec that decomposes input audio into separate frequency bands, each of which is processed independently and in parallel through dedicated encoder, quantizer, and decoder modules. While band-splitting provides a physically grounded design philosophy, its effective application to codec design requires careful consideration of band configuration and quantization allocation. We conduct extensive experiments to systematically investigate the impact of band count, frequency boundaries, and codebook design on multi-domain codec performance. Through principled design of band partitioning and quantization structure, our approach achieves strong performance across all three domains. Specifically, under the same training conditions, our 2.55 kbps and 3.83 kbps models achieve comparable overall performance to DAC at 4.5 kbps and 6 kbps respectively.

Our contributions can be summarized as follows:
\begin{itemize}
    \item We propose BSCodec, a band-split codec that independently processes time-domain signals from separated frequency bands through parallel encoder-quantizer-decoder modules, achieving strong multi-domain performance through carefully designed band configuration and quantization allocation.
    
    \item Comprehensive experiments on reconstruction quality and downstream tasks demonstrate that BSCodec achieves both perceptual superiority and enhanced effectiveness for audio understanding.

\end{itemize}

\section{Related Work}
\label{sec:rel}
\noindent\textbf{Neural Audio Codecs}.
% Neural Audio Codecs (NACs) are designed to compress continuous audio signals into a compact sequence of discrete tokens, aiming for high-quality reconstruction with maximum efficiency in storage and transmission.
% Neural audio codecs (NACs) are designed to compress continuous audio waveforms into compact sequences of discrete tokens while preserving perceptual quality and maximizing storage and transmission efficiency. The dominant paradigm for NACs builds upon the Vector Quantized Variational Autoencoder (VQ-VAE) framework, which comprises three principal components: an encoder that projects input waveforms into a continuous latent space, a vector quantizer that discretizes these representations through nearest-neighbor mapping to learned codebooks, and a decoder that synthesizes audio from the resulting discrete tokens.
The advent of deep learning has enabled significant advances in neural audio compression. They typically adopt the core architecture of RVQ-based discretization paired with GAN~\cite{goodfellow2014generative}-driven reconstruction, and further improve performance via targeted architectural enhancements or optimized training strategies~\cite{zeghidour2021soundstream, defossez2022high, kumar2023high}.

% SoundStream~\cite{zeghidour2021soundstream} proposed a neural codec incorporating Residual Vector Quantization (RVQ) to enhance quantization capacity and Generative Adversarial Networks (GANs)~\cite{goodfellow2014generative} to improve reconstruction quality. Encodec~\cite{defossez2022high} advanced this approach by integrating multi-scale STFT discriminators and a loss-balancing strategy, establishing new benchmarks for perceptual quality. Descript Audio Codec (DAC)~\cite{kumar2023high} further refined RVQ-based architectures through architectural innovations and training strategies, consolidating the residual quantization paradigm as the dominant approach for neural audio compression.

Using a unified model remains challenging due to the distinct energy distribution of universal audio. Recent efforts have pursued universal codecs capable of handling diverse audio types. However, achieving strong multi-domain performance remains challenging due to the distinct acoustic characteristics across domains—speech exhibits narrow-band harmonic structure, music requires wide-band spectral richness, and sound encompasses diverse sound textures~\cite{scharf1970critical}. Existing multi-domain codecs address this heterogeneity through various specialized mechanisms. DAC~\cite{kumar2023high} employs balanced sampling strategies, ensuring each training batch contains data from all three domains (speech, music, sound) to prevent domain bias. 
Further works~\cite{jiang2025unicodec, liu2024semanticodec, yang2025almtokenizer} introduce domain-specific codebooks via Mixture-of-Experts layers~\cite{MoE}, semantic priors and MAE-derived representations~\cite{MAE, huang2022masked}, and more fine-grained supervision to further enhance the language modeling capability of universal audio codecs.
% Unicodec~\cite{jiang2025unicodec} introduces domain-conditioned quantization through Mixture-of-Experts (MoE) layers in the encoder and domain-specific codebook entries, where domain labels explicitly control which codebook vectors are updated during training. SemantiCodec~\cite{liu2024semanticodec} leverages pre-trained AudioMAE~\cite{huang2022masked} representations to inject semantic information for general sound reconstruction. While effective for downstream semantic tasks, these models still lag behind RVQ-based universal codecs in reconstruction quality, and their substantial bitrate differences complicate fair comparison.

In contrast, our work adopts a simpler design philosophy. We employ an architecture and training framework closely aligned with DAC~\cite{kumar2023high}, yet achieve comparable performance for both reconstruction and downstream tasks at half the bitrate on the same dataset. This demonstrates that band-split decomposition offers a simple yet effective alternative for multi-domain codec design, obviating the need for complex domain-specific mechanisms while maintaining strong generalization across audio types.
\begin{figure*}[t]
  \centering
  \includegraphics[width=\textwidth]{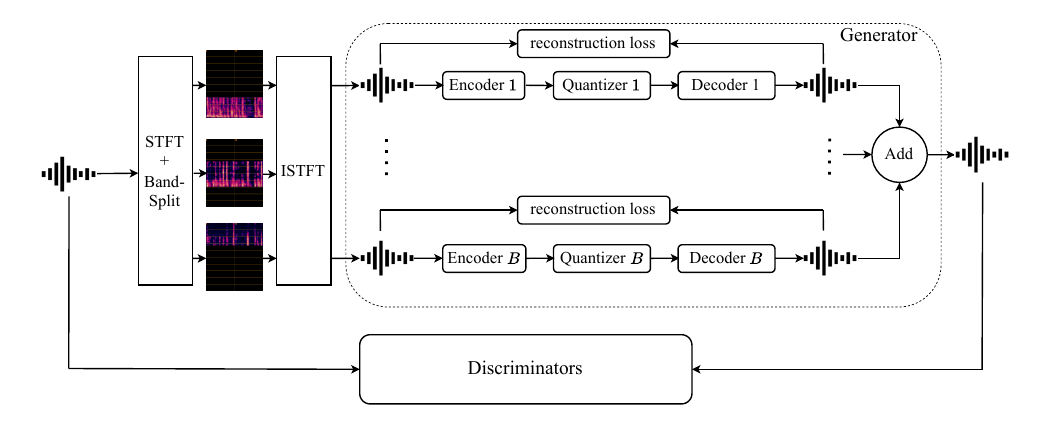}
  \caption{BSCodec architecture with band split, multi-band parallel generators and discriminators.}
  \label{fig:wide_architecture}
  \vspace{-10pt}
\end{figure*}

\noindent\textbf{Band-Splitting for Audio Processing}.
% Frequency-domain decomposition of audio signals into distinct sub-bands is a fundamental principle in audio processing, motivated by the observation that different spectral regions exhibit distinct perceptual and structural characteristics that benefit from independent modeling.
% In classical signal processing, band decomposition is realized through filter banks (e.g., QMF) or time-frequency transforms (e.g., STFT, wavelet transforms), forming the foundation for perceptual audio coding standards such as MP3 and AAC. This principle has been increasingly adopted in deep learning. Conv-TasNet introduced learnable filter banks for source separation, demonstrating that adaptive band decomposition can outperform hand-crafted transforms. In neural audio synthesis, multi-band vocoders like HiFi-GAN employ separate branches for low and high frequencies, exploiting their different acoustic properties—low frequencies encode harmonic structure while high frequencies convey texture and transients.
% Recent work has applied band-splitting to neural audio coding. Split-band autoencoders allocate different bitrates to frequency bands based on perceptual importance. However, these approaches primarily focus on signal-level processing or generation. In contrast, our work applies frequency decomposition directly to the quantization structure, organizing multiple codebook layers according to acoustically meaningful spectral partitions rather than residual hierarchies.
Band-splitting decomposes audio signals into separate frequency bands for independent processing, leveraging the observation that different spectral regions carry distinct perceptual and structural information: low frequencies encode harmonic structure and fundamental pitch, mid frequencies capture formants and timbral characteristics, while high frequencies convey transients and texture. This principle has deep roots in classical signal processing. Subband decomposition techniques such as QMF~\cite{malvar1990modulated} and polyphase filter banks~\cite{saramaki2002multirate}, combined with time-frequency transforms like 
MDCT~\cite{prokop2003general}, form the foundation for perceptual audio coding 
standards like MP3~\cite{brandenburg1999mp3} and AAC. Wavelet transforms~\cite{zhang2019wavelet} similarly provide multi-resolution frequency decomposition.

Recent deep learning approaches have demonstrated the effectiveness of band-split architectures across various audio tasks. In music source separation, Band-split RNN~\cite{luo2023music} partitions the spectrogram into multiple frequency bands, processing each band with dedicated recurrent networks before reconstruction. This frequency-aligned decomposition enables the model to specialize in acoustically coherent spectral regions, achieving superior separation quality compared to full-band processing. In neural audio synthesis, Multi-band MelGAN~\cite{yang2021multi} employs separate generator branches for different frequency bands, significantly improving synthesis quality and efficiency by exploiting the distinct acoustic properties of each spectral region.

These successes demonstrate that architectures aligned with the natural spectral structure of audio yield improved performance, providing insights for the design of BSCodec. 

\section{Methodology}
\label{sec:met}
The proposed BSCodec follows an encoder-quantizer-decoder architecture with adversarial training, adopting a framework similar to DAC~\cite{kumar2023high}. The overall architecture is illustrated in Figure \ref{fig:wide_architecture}.

\subsection{Band Splitting}

The model first decomposes the input audio waveform into several band-limited signals through time-frequency domain processing. The input discrete-time signal $x[n]$ is transformed into a spectrogram $X(m, k)$ using the Short-Time Fourier Transform~(STFT):
\begin{equation}
X(m, k) = \sum_{n=0}^{N-1} x[n+mR] w[n] e^{-j2\pi kn/N}
\end{equation}
where $N$ denotes the window length (FFT size), $n$ is the sample index within the window, $w[n]$ denotes the window function, $R$ the hop size, $m$ the frame index, and $k$ the frequency bin index.

The band-splitting operation is then performed directly on $X(m, k)$. For an input sampling rate $f_s = 24$~kHz, we define $B$ non-overlapping frequency bands with boundaries $\{f_0, f_1, \dots, f_B\}$. Each band $b$ is isolated using a binary mask $M_b(k)$:
\begin{equation}
M_b(k) = \begin{cases} 1 & \text{if } f_{b-1} \le \frac{k \cdot f_s}{N} < f_b \\ 0 & \text{otherwise} \end{cases}
\end{equation}
The band-specific spectrogram $X_b(m, k)$ is obtained via element-wise multiplication: $X_b(m, k) = X(m, k) \odot M_b(k)$. Each masked spectrogram is then transformed back to the time domain via Inverse STFT:
% \begin{multline}
% x_b[n] = \sum_{m} w[n-mR] \times \\
% \left( \frac{1}{N} \sum_{k=0}^{N-1} X_b(m, k) e^{j2\pi kn/N} \right)
% \end{multline}
\begin{equation}
\resizebox{\linewidth}{!}{$
x_b[n] = \sum_{m} w[n - mR]\times\left(\frac{1}{N} 
\sum_{k=0}^{N-1} X_b(m, k) e^{j 2\pi k n / N}\right)
$}
\end{equation}

\subsection{Encoder and Decoder}
% In the encoding stage, the input audio is first separated into band-limited waveforms using the STFT/ISTFT process described previously.

% The band-splitting process operates in the frequency domain on the complex spectrogram $X(m, k)$. We first define a set of $B$ non-overlapping frequency intervals, such as $(f_0, f_1), (f_1, f_2), \dots, (f_{B-1}, f_B)$.

% For each band $b$, a binary mask $M_b(k)$ is created and applied across the frequency axis ($k$). The mask's value is 1 for frequency bins falling within the band's range and 0 otherwise. This can be expressed as:

% $$M_b(k) = \begin{cases} 1 & \text{if } f_{b-1} \le \frac{k \cdot \text{sr}}{N} < f_b \\ 0 & \text{otherwise} \end{cases}$$

% Here, $\text{sr}$ is the sampling rate and $N$ is the FFT size. The spectrogram for each band, $X_b(m, k)$, is then isolated by performing an element-wise multiplication of the full spectrogram with the corresponding mask:

% $$X_b(m, k) = X(m, k) \odot M_b(k)$$

% Finally, a separate ISTFT is applied to each of these masked spectrograms, $\{X_1, X_2, \dots, X_B\}$, to yield the set of band-limited waveforms that are fed into the parallel encoders.

The resulting band-limited waveforms $\{x_1[n], x_2[n], \dots, x_B[n]\}$ are processed in parallel by independent encoders. Each encoder adopts the SEANet~\cite{tagliasacchi2020seanet} architecture with downsampling strides of $[2, 4, 5, 8]$ across four stages. Starting from an initial channel dimension of 32 that doubles at each stage, the encoders produce latent features with 512 dimensions at a frame rate of 75~Hz. Each convolutional block contains 3 residual units followed by strided convolution. Critically, no parameters are shared across the $B$ encoders, allowing each to specialize in the spectral characteristics of its frequency range.

The decoder architecture mirrors the encoder with symmetric upsampling strides of $[8, 5, 4, 2]$, progressively upsampling the quantized representations from 512 dimensions back to 24~kHz waveforms. The final output $\hat{x}[n]$ is obtained by summing all band-specific reconstructions: $\hat{x}[n] = \sum_{b=1}^{B} \hat{x}_b[n]$.

\subsection{Vecter Quantization}
The vector quantization stage discretizes the continuous latent representations from each band's encoder into discrete tokens. We employ a single-layer quantization scheme per band, utilizing SimVQ\cite{zhu2024addressing} to accommodate configurations requiring large codebook capacities.

SimVQ\cite{zhu2024addressing} enhances the standard Vector Quantization (VQ) framework by introducing a learnable linear transformation on the codebook embeddings. Given the encoder output $z$ and a codebook of size $K$: $\{\mathbf{c}_1, \mathbf{c}_2, \ldots, \mathbf{c}_K\}$, the quantization proceeds as follows. First, the nearest codebook entry is determined by:
\begin{equation}
q = \arg\min_{\mathbf{c} \in \{\mathbf{c}_1, \ldots, \mathbf{c}_K\}} \|z - \mathbf{c}W\|
\end{equation}
where $W$ is a learnable transformation matrix applied to the codebook embeddings. The quantized representation is then computed as:
\begin{equation}
z_q = z + \text{sg}[qW - z]
\end{equation}
where $\text{sg}[\cdot]$ denotes the stop-gradient operation. During forward propagation, this formulation passes the transformed codebook entry $qW$ to the decoder, while during backpropagation, gradients flow directly through $z$ via the straight-through estimator.

The codebook is optimized through a bidirectional commitment loss with a hyperparameter $\lambda$:
\begin{equation}
\mathcal{L}_{\text{commit}} = \|\text{sg}[z] - qW\|^2 + \lambda \|z - \text{sg}[qW]\|^2
\label{eq:commitment}
\end{equation}
The first term encourages the quantized codes to stay close to the encoder outputs, while the second term allows the encoder to adapt towards the codebook. The transformation matrix $W$ improves optimization dynamics by providing a reparameterization that facilitates gradient-based learning of large codebooks. In our experiments, each band is assigned a SimVQ codebook of size $K = 131{,}072$.

\subsection{Loss Design}

\noindent\textbf{Reconstruction Loss.} We adopt a multi-scale mel-spectrogram L1 loss for frequency-domain reconstruction. Specifically, we compute mel-spectrograms at multiple scales ranging from $2^6$ to $2^{11}$ using a Hann window with 80 mel-frequency bins. To provide better supervision signals for each band, we additionally compute the mel loss between each individual band's reconstructed audio and its corresponding ground-truth band, then average these losses and incorporate them into the overall optimization objective.

\noindent\textbf{Adversarial Loss.} Following DAC~\cite{kumar2023high}, we employ a multi-period discriminator for waveform-level discrimination and a multi-band multi-scale STFT discriminator for frequency-domain discrimination, using the hinge loss formulation with feature matching loss.

\noindent\textbf{Codebook Learning.} The codebook commitment loss follows Equation~\eqref{eq:commitment} with $\lambda$ set to 0.25. Gradients are backpropagated through the codebook lookup using the straight-through estimator.

% \noindent\textbf{Loss weighting}
\noindent\textbf{Loss Weighting.} We set the loss weighting coefficients as follows: 45.0 for the multi-scale mel loss, 2.0 for the feature matching loss, 1.0 for both the adversarial loss and reconstruction loss, and 1.0 for the codebook commitment loss.

% \subsection{Training objective}
% \textbf{Reconstruction Loss} Our reconstruction loss is composed of two key components. The primary component is the multi-scale mel-spectrogram loss from DAC, which measures the L1 distance in the spectral domain at multiple scales. This loss is effective as the mel scale directly relates to perceptual quality, making it preferable to the multi-scale STFT loss. Alongside this spectral objective, we also incorporate a time-domain L1 reconstruction loss.

% \textbf{Reconstruction Loss}
% \textbf{GAN loss} We use the loss function of least-square GAN to train the model, which uses least square loss instead of conventional
% binary cross entropy loss to stabilize training. Additionally, the L1
% feature matching loss proposed in HiFi-GAN is incorporated.

\section{Experiments}
\label{sec:exp}
\subsection{Experimental Stages}
We conduct our experiments in two progressive stages to systematically validate the effectiveness of band split across different domains.

\noindent\textbf{Stage 1: Music Domain Validation.} We first validate the feasibility of band split on music domain. Music is chosen as the initial testbed because its wider fundamental frequency distribution makes it a less challenging learning task for the model. At this stage, we explore various configurations of band split. The experimental results demonstrate that band split achieves excellent performance on music, establishing a strong foundation for further investigation.

\noindent\textbf{Stage 2: Multi-Domain Extension.} Building on the success in music, we extend our investigation to a multi-domain setting encompassing speech, music and sound. We aim to determine whether band split maintains its effectiveness when dealing with diverse acoustic characteristics across domains. Our experiments show that band split remains effective in the multi-domain scenario. Furthermore, we introduce targeted optimizations to enhance model performance across all domains.

\subsection{Datasets}
In Stage 1, we train on Jamendo\cite{bogdanov2019mtg}, MUSDB18\cite{musdb18}, and MAESTRO\cite{hawthorne2018enabling}. We evaluate on two test sets: the MUSDB18 test set for vocal music, and 100 clips from MAESTRO for non-vocal music.

In Stage 2, we train on approximately 2,100 hours of data spanning three domains. For speech, we use LibriTTS~\cite{zen2019libritts}, VCTK~\cite{yamagishi2019vctk}, and CommonVoice~\cite{ardila2019common} as training data, evaluating on the LibriTTS test-clean. For music, we train on Jamendo\cite{bogdanov2019mtg} and MUSDB18\cite{musdb18}, evaluating on the MUSDB18 test set. For sound, we train on AudioSet\cite{gemmeke2017audio} and evaluate on the AudioSet test set. To ensure domain balance, We sample about 700 hours from each domain for training. The detailed dataset distribution after sampling is shown in the Appendix~\ref{app:4}.

% \begin{table*}[t]
% \centering
% \caption{Performance comparison on music and instrument domains}
% \label{tab:music_instrument_results}
% \begin{tabular}{lccccccc}
% \toprule
% \multicolumn{2}{c}{Model} & \multicolumn{3}{c}{Music} & \multicolumn{3}{c}{Instrument} \\
% \cmidrule(lr){1-2} \cmidrule(lr){3-5} \cmidrule(lr){6-8}
% Model & Bitrate & VISQOL$\uparrow$ & Mel Dist.$\downarrow$ & STFT Dist.$\downarrow$ & VISQOL$\uparrow$ & Mel Dist.$\downarrow$ & STFT Dist.$\downarrow$ \\
% \midrule
% DAC & 6 kbps & 4.097 & 0.481 & 1.018 & 4.479 & 0.506 & 0.959 \\
% 5-band VQ & 3.75 kbps & \textbf{4.384} & \textbf{0.445} & \textbf{0.921} & 4.470 & 0.419 & 0.788 \\
% 5-band SimVQ & 3.75 kbps & 4.274 & 0.493 & 1.070 & 4.484 & 0.442 & 0.841 \\
% 3-band SimVQ & 3.825 kbps & 4.196 & 0.477 & 1.042 & 4.495 & \textbf{0.412} & \textbf{0.787} \\
% 2-band SimVQ & 2.55 kbps & 4.104 & 0.482 & 0.994 & 4.484 & 0.417 & 0.806 \\
% RSimVQ & 4.5 kbps & 4.106 & 0.475 & 1.039 & \textbf{4.514} & 0.444 & 0.848 \\
% \bottomrule
% \end{tabular}
% \end{table*}

\begin{table*}[h]
\centering
\small
\setlength{\tabcolsep}{4pt}
\caption{Reconstruction performance of BSCodec and baseline DAC on vocal songs and instrumental music.}
\vspace{-10pt}
\label{tab:music_instrument_results}
\begin{tabular}{lllcccccc}
\toprule
\multicolumn{3}{c}{Model} & \multicolumn{3}{c}{Vocal songs} & \multicolumn{3}{c}{Instrumental music} \\
\cmidrule(lr){1-3} \cmidrule(lr){4-6} \cmidrule(lr){7-9}
Codec & VQ Type & Bitrate & VISQOL$\uparrow$ & Mel Dist.$\downarrow$ & STFT Dist.$\downarrow$ & VISQOL$\uparrow$ & Mel Dist.$\downarrow$ & STFT Dist.$\downarrow$ \\
\midrule
DAC & RVQ & 6.00 kbps & 4.097 & 0.481 & 1.018 & 4.479 & 0.506 & 0.959 \\
DAC & RVQ & 4.50 kbps & 4.075 & 0.493 & 1.025 & 4.459 & 0.517 & 0.962
\\
DAC & RVQ & 3.00 kbps & 4.030 & 0.515 & 1.039 & 4.423 & 0.541 & 0.968
\\
BSCodec & 5-band VQ & 3.75 kbps & \textbf{4.384} & \textbf{0.445} & \textbf{0.921} & 4.470 & 0.419 & 0.788 \\
% BSCodec & 5-band SimVQ & 3.75 kbps & 4.274 & 0.493 & 1.070 & 4.484 & 0.442 & 0.841 \\
BSCodec & 3-band SimVQ & 3.83 kbps & 4.196 & 0.477 & 1.042 & \textbf{4.495} & \textbf{0.412} & \textbf{0.787} \\
BSCodec & 2-band SimVQ & 2.55 kbps & 4.104 & 0.482 & 0.994 & 4.484 & 0.417 & 0.806 \\
% RVQ-Codec(?) & RSimVQ & 4.50 kbps & 4.106 & 0.475 & 1.039 & \textbf{4.514} & 0.444 & 0.848 \\
\bottomrule
\end{tabular}
\vspace{-10pt}
\end{table*}

% \subsection{Training details}
% Our model was built and trained based on the ESPnet framework. The training was conducted with a global batch size of 72, where all audio data was processed into 1-second chunks (corresponding to 24,000 samples). For optimization, we utilized the AdamW optimizer with an initial learning rate of $2.0 \times 10^{-4}$ and betas $(\beta_1, \beta_2)$ set to $(0.5, 0.9)$. This was coupled with an exponential learning rate scheduler, which applied a decay factor of $\gamma = 0.999875$ during training. The total number of training iterations is $340k$.

% \subsection{Band Split}
% We experimented with splitting the signal into 5, 3, and 2 frequency bands. The 5-band configuration splits the signal into $[0, 0.5]$, $[0.5, 2]$, $[2, 4]$, $[4, 8]$, and $[8, 12]$ kHz. For coarser divisions, we tested a 3-band split ($[0, 2]$, $[2, 4]$, $[4, 12]$ kHz) and a 2-band split ($[0, 2]$, $[2, 12]$ kHz). Since configurations with fewer bands require a larger codebook size for each sub-band to maintain quality, we selected SimVQ as our quantization method.

\subsection{Training Setup}
\label{sec:training_details}
Our model is implemented using the codec part~\cite{shi2024espnet} in ESPnet~\cite{watanabe2018espnet} toolkit. We train with a global batch size of 72, processing audio into 1-second chunks at a sampling rate of 24 kHz. We employ the AdamW~\cite{loshchilov2017decoupled} optimizer with an initial learning rate of $2.0 \times 10^{-4}$ and momentum coefficients $(\beta_1, \beta_2) = (0.5, 0.9)$. The learning rate is decayed exponentially with a factor of $\gamma = 0.999875$ per epoch, where each epoch processes 2000 samples (approximately 1200 iterations). Training proceeds for 340k iterations in total.
For comparison, we train a DAC baseline with 8 codebook layers using identical training configurations and datasets. We use the resulting 6 kbps model, along with its 4.5 kbps and 3 kbps variants obtained via codebook dropout, as baselines for comparison.

\subsection{Band Partitioning Configurations}
We investigate band partitioning strategies with 5, 3, and 2 frequency bands. The 5-band configuration partitions the spectrum into $[0, 0.5]$, $[0.5, 2]$, $[2, 4]$, $[4, 8]$, and $[8, 12]$ kHz. For coarser granularities, we evaluate a 3-band configuration ($[0, 2]$, $[2, 4]$, $[4, 12]$ kHz) and a 2-band configuration ($[0, 2]$, $[2, 12]$ kHz). Our partitioning strategy is inspired by the frequency division in band-split RNN~\cite{luo2023music}. While Band-Split RNN uses 41 bands, we adopt only the density distribution of its partitioning. As configurations with fewer bands necessitate larger codebook sizes per sub-band to maintain overall model capacity, we adopt SimVQ\cite{zhu2024addressing} for quantization, which efficiently handles large codebooks.

\subsection{Evaluation Metrics}

We evaluate our model using domain-specific metrics. For speech, we use Mel Cepstral Distortion (MCD), WB-PESQ~\cite{rix2001perceptual}, STOI~\cite{taal2010short}, Speaker Similarity (SPK\_SIM)\cite{jung2024espnet} and UTMOS~\cite{saeki2022utmos}. For audio and music, we use VISQOL\cite{hines2015visqol}, Mel Distance and STFT Distance. All metrics except Mel Distance and STFT Distance are evaluated using the VERSA~\cite{shi2024versa} toolkit with default configurations.

\noindent\textbf{Mel Distance} is computed as the L1 distance between mel-scaled magnitude spectrograms using multi-resolution STFT with Hann window lengths ranging from $2^6$ to $2^{11}$ samples projected onto an 80-bin Mel filterbank.

\noindent\textbf{STFT Distance} is computed as the L1 distance using multi-scale STFT with a 2048-sample window and 512-sample hop as well as a 512-sample window and 128-sample hop.

\section{Results and Discussions}
\label{sec:res}

\subsection{Music Reconstruction}
\label{sec:musrec}

We first evaluate on the music domain, which exhibits a broad fundamental frequency distribution. We compare against the DAC baselines described in Section~\ref{sec:training_details}. We assess reconstruction quality on vocal music and solo piano datasets using VISQOL, Mel Distance, and STFT Distance metrics.

As shown in Table~\ref{tab:music_instrument_results}, our 5-band VQ configuration achieves the best performance on music at 3.75 kbps compared to both DAC baselines, demonstrating that fine-grained frequency decomposition effectively captures harmonic content. On the instrument domain, the 3-band SimVQ achieves competitive results at 3.83 kbps, while the 2-band configuration maintains strong performance at only 2.55 kbps. These results demonstrate that band partitioning is highly effective for domains with wide frequency distributions, with optimal granularity varying by domain complexity.

\begin{table*}[t]
\centering
\scriptsize
\setlength{\tabcolsep}{2pt}
\caption{Reconstruction performance on speech, sound and music domains. $\dagger$ means the official release.}
\vspace{-10pt}
\label{tab:reconstruction_results}
\begin{tabular}{lllccccccccccc}
\toprule
\multicolumn{3}{c}{Model} & \multicolumn{5}{c}{Speech} & \multicolumn{3}{c}{Sound} & \multicolumn{3}{c}{Music} \\
\cmidrule(lr){1-3} \cmidrule(lr){4-8} \cmidrule(lr){9-11} \cmidrule(lr){12-14}
Codec & VQ Method & Bitrate & MCD$\downarrow$ & PESQ$\uparrow$ & STOI$\uparrow$ & SPK\_SIM$\uparrow$ & UTMOS$\uparrow$ & VISQOL$\uparrow$ & Mel Dist.$\downarrow$ & STFT Dist.$\downarrow$ & VISQOL$\uparrow$ & Mel Dist.$\downarrow$ & STFT Dist.$\downarrow$ \\
\midrule
EnCodec$^\dagger$ & RVQ & 6.00 kbps & 5.94 & 2.715 & 0.939 & 0.865 & 3.038 & 4.240 & 0.485 & 0.940 & 4.410 & 0.435 & 0.980 \\
DAC & RVQ & 6.00 kbps & 5.40 & 2.915 & 0.934 & 0.751 & 3.356 & 4.085 & 0.452 & 0.874 & 4.201 & 0.439 & 0.974 \\
\midrule
DAC & RVQ & 4.50 kbps & 5.50 & \textbf{2.726} & \textbf{0.925} & 0.734 & 3.201 & 4.055 & 0.463 & 0.880 & 4.171 & \textbf{0.449} & 0.979 \\
DAC & RVQ & 3.00 kbps & 5.74 & 2.397 & 0.905 & 0.686 & 2.869 & 3.990 & 0.485 & 0.893 & 4.105 & 0.472 & 0.993 \\
EnCodec$^\dagger$ & RVQ & 3.00 kbps & 6.49 & 2.048 & 0.901 & 0.771 & 2.305 & 4.085 & 0.531 & 0.978 & 4.262 & 0.481 & 1.014 \\
BSCodec & 5\# VQ & 3.75 kbps & 5.08 & 1.961 & 0.894 & 0.810 & 2.515 & \textbf{4.245} & 0.463 & 0.800 & \textbf{4.326} & 0.464 & 0.892 \\
BSCodec & 3\# SimVQ & 3.83 kbps & \textbf{5.05} & 2.544 & 0.920 & \textbf{0.852} & \textbf{3.360} & 4.234 & \textbf{0.456} & \textbf{0.794} & 4.298 & 0.461 & \textbf{0.888} \\
BSCodec & 2\# SimVQ & 2.55 kbps & 5.42 & 2.429 & 0.916 & 0.783 & 3.304 & 4.137 & 0.470 & 0.846 & 4.166 & 0.479 & 0.916 \\
\bottomrule
\end{tabular}
\vspace{-10pt}
\raggedright
\end{table*}
% \begin{table*}[t]
% \centering
% \small
% \setlength{\tabcolsep}{4pt}
% \caption{Performance comparison on audio and music domains}
% \label{tab:audio_music_results}
% \begin{tabular}{lllcccccc}
% \toprule
% \multicolumn{3}{c}{Model} & \multicolumn{3}{c}{Audio} & \multicolumn{3}{c}{Music} \\
% \cmidrule(lr){1-3} \cmidrule(lr){4-6} \cmidrule(lr){7-9}
% Codec & VQ Type & Bitrate & VISQOL$\uparrow$ & Mel Dist.$\downarrow$ & STFT Dist.$\downarrow$ & VISQOL$\uparrow$ & Mel Dist.$\downarrow$ & STFT Dist.$\downarrow$ \\
% \midrule
% DAC & RVQ & 6.00 kbps & 4.085 & \textbf{0.452} & 0.874 & 4.201 & \textbf{0.439} & 0.974 \\
% DAC & RVQ & 4.50 kbps & 4.055 & 0.463 & 0.880 & 4.171 & 0.449 & 0.979 \\
% % RVQ-Codec(?) & RSimVQ & 4.50 kbps & 4.068 & 0.469 & 0.851 & 4.172 & 0.444 & 0.899 \\
% BSCodec & 5-band VQ & 3.75 kbps & \textbf{4.245} & 0.463 & 0.800 & \textbf{4.326} & 0.464 & 0.892 \\
% % BSCodec & 5-band SimVQ & 3.75 kbps & 4.185 & 0.476 & 0.814 & 4.267 & 0.476 & 0.903 \\
% BSCodec & 3-band SimVQ & 3.83 kbps & 4.234 & 0.456 & \textbf{0.794} & 4.298 & 0.461 & \textbf{0.888} \\
% BSCodec & 2-band SimVQ & 2.55 kbps & 4.137 & 0.470 & 0.846 & 4.166 & 0.479 & 0.916 \\
% \bottomrule
% \end{tabular}
% \end{table*}

\subsection{Multi-domain Reconstruction}
\label{sec:mdrec}

We conduct experiments on a strictly balanced dataset comprising three domains: speech, audio, and music. We train our model alongside a DAC baseline using identical configurations. For speech evaluation, we employ MCD, WB-PESQ, STOI, SPK\_SIM and UTMOS as metrics. For audio and music domains, we utilize VISQOL, Mel Distance and STFT Distance.

As shown in Table~\ref{tab:reconstruction_results}, the speech domain, characterized by a narrow and concentrated fundamental frequency range, poses significant challenges for band splitting. While the 5-band VQ configuration achieves excellent results on audio and music domains, it yields suboptimal performance on speech. Although MCD remains competitive, perceptually-oriented metrics reveal substantial limitations compared to the DAC baseline.

To address this challenge, we adopt a coarser partitioning strategy for speech. Rather than forcing fine-grained low-frequency decomposition, we employ the 3-band SimVQ configuration at 3.83 kbps, which uses a larger codebook to handle a unified low-frequency region while maintaining high-frequency splitting. This modification achieves substantial improvements on speech across all perceptual metrics, with speaker similarity significantly outperforming the DAC baseline. The enhanced speaker similarity demonstrates that our approach effectively preserves speaker-specific characteristics, particularly through the improved low-frequency representation combined with accurate high-frequency reconstruction of formants, fricatives, and plosives.

Notably, the 3-band configuration maintains competitive performance on audio and music domains, with virtually no degradation compared to the 5-band variant. This demonstrates that our adaptive partitioning strategy successfully balances the trade-off between speech-specific requirements and multi-domain effectiveness. The 2-band configuration further reduces bitrate to 2.55 kbps while maintaining reasonable performance across all domains, offering an efficient alternative for bandwidth-constrained scenarios.

\begin{table*}[t]
\centering
\small
\setlength{\tabcolsep}{4pt}
\renewcommand{\arraystretch}{0.95}
\caption{Performance comparison on ARCH.}
\vspace{-10pt}
\label{tab:classification_results_2}
\begin{tabular}{llccccccc}
\toprule
\multicolumn{2}{c}{Model} & \multicolumn{2}{c}{Speech} & \multicolumn{3}{c}{Audio} & \multicolumn{2}{c}{Music} \\
\cmidrule(lr){1-2} \cmidrule(lr){3-4} \cmidrule(lr){5-7} \cmidrule(lr){8-9}
Model & \# Enc. & RAVDESS$\uparrow$ & AM$\uparrow$ & ESC-50$\uparrow$ & US8K$\uparrow$ & VIVAE$\uparrow$ & MTT$\uparrow$ & MS-DB$\uparrow$ 
\\
\midrule
DAC & 1 & 0.3958 & 0.7791 & 0.3335 & 0.5311 & 0.3285 & 0.2949 & 0.5754 \\
BSCodec & 2 & 0.4201 & 0.7801 & 0.3795 & 0.5798 & 0.3265 & \textbf{0.3555} & \textbf{0.7021} 
\\
BSCodec & 3 & 0.4306 & 0.7759 & 0.3725 & 0.5654 & 0.3258 & 0.3531 & 0.6969 
\\
BSCodec & 5 & \textbf{0.5069} & \textbf{0.8548} & \textbf{0.3930} & \textbf{0.5956} & \textbf{0.3810} & 0.2580 & 0.5751
\\
\bottomrule
\end{tabular}
\vspace{-10pt}
\end{table*}

% As shown in Figure~\ref{fig:codebook_analysis}, despite achieving significantly better reconstruction quality than DAC, our model maintains comparable overall codebook utilization (98.63\% for DAC vs. 92.84\% for BSCodec-3band). We further analyze the joint utilization of adjacent codebook layers following the methodology described in the method section. The results reveal that the joint utilization of DAC's first two layers is comparable to our large single codebook. For DAC, as residual layers progress, the encoded information approaches random noise, resulting in increasingly high utilization rates.

\subsection{Codebook Utilization}

\begin{figure}[t]
\centering
\includegraphics[width=\columnwidth]{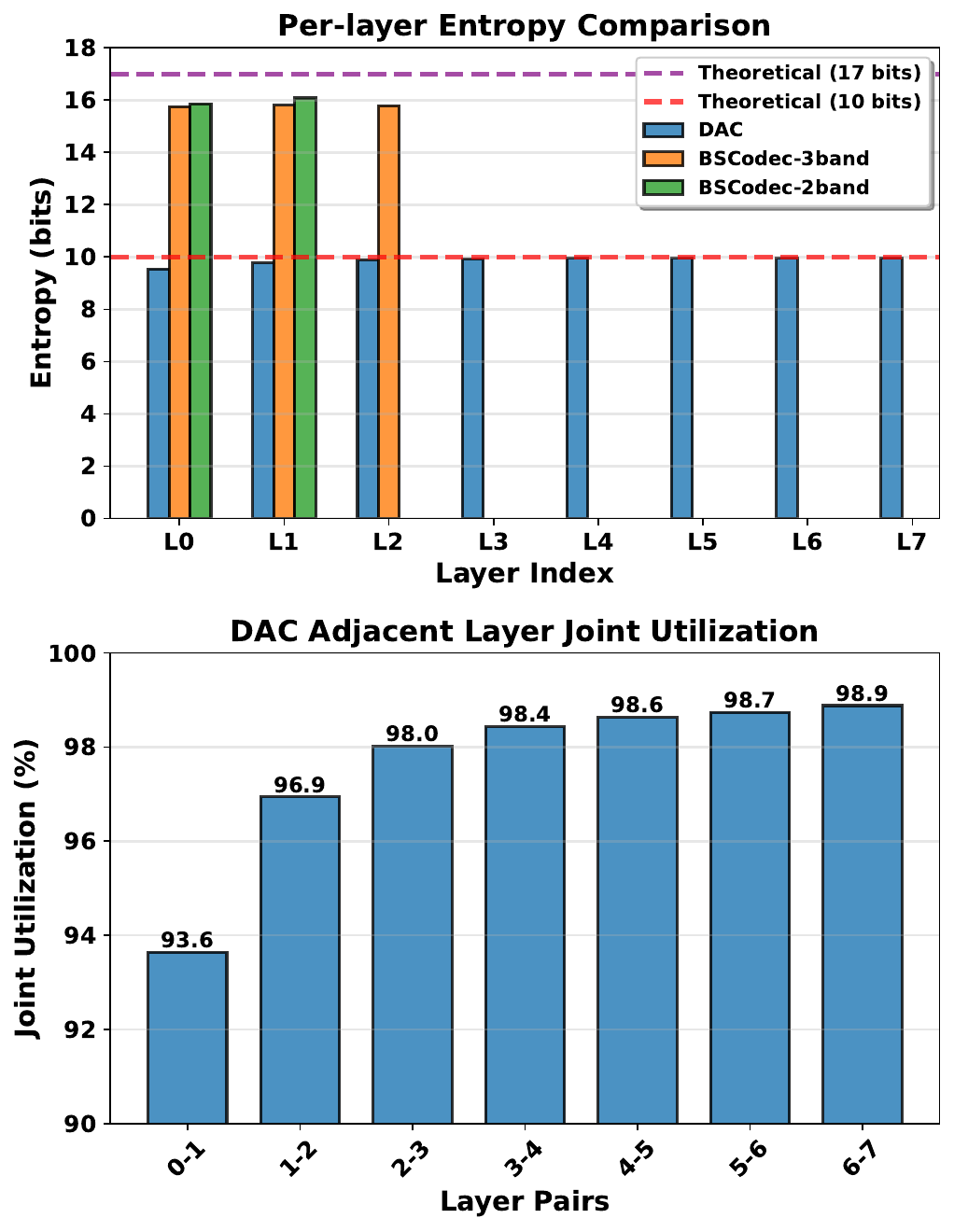}
\caption{Codebook utilization analysis. Top: Per-layer entropy comparison across different models. Bottom: Joint utilization of adjacent layers in DAC.}
\label{fig:codebook_analysis}
\vspace{-10pt}
\end{figure}

We evaluate codebook utilization on models trained across three domains. We collect statistics on codebook usage frequencies using a test set comprising LibriTTS test-clean and test-other, AudioSet test set, and MUSDB18 test set, totaling 237k seconds of audio data. As shown in Figure~\ref{fig:codebook_analysis}, when examining per-layer utilization, our model exhibits slightly lower rates than DAC (98.63\% for DAC vs. 92.84\% for BSCodec-3band). To analyze inter-codebook correlations, we compute joint utilization by combining codewords from two adjacent codebooks into a larger codebook. The calculation methods for single-layer and joint codebook utilization are detailed in Appendix~\ref{app:2}. The joint utilization for each pair of adjacent codebooks is shown in Figure~\ref{fig:codebook_analysis}. The layer-wise growth pattern remains consistent with single-layer observations, but notably, DAC's first two layers exhibit strong correlation, achieving the same joint utilization to our single-layer SimVQ. Moreover, DAC's final two layers show very high joint utilization. However, reconstruction results demonstrate that DAC's final two layers, despite exhibiting very high joint utilization (totaling 20 bits), contribute minimally to performance improvement. In contrast, when we reduce our model from 3-band to 2-band by removing the 17-bit codebook in the high-frequency region, speech performance degrades only slightly at 2.55 kbps, but notable performance gaps emerge in both audio and music domains compared to the 3-band configuration. This suggests that our band-specific codebooks capture domain-critical information more effectively than residual quantization approaches.

% However, reconstruction results demonstrate that DAC's final two layers, despite exhibiting very high joint utilization (totaling 20 bits), contribute minimally to performance improvement. In contrast, when we reduce our model from 3-band to 2-band by removing the 17-bit codebook in the high-frequency region, speech performance degrades only slightly at 2.55 kbps, but notable performance gaps emerge in both audio and music domains compared to the 3-band configuration. This suggests that our band-specific codebooks capture domain-critical information more effectively than residual quantization approaches.

% \begin{table*}[htbp]
% \centering
% \caption{Performance on downstream tasks}
% \label{tab:downstream_results}
% \begin{tabular}{lllcccc}
% \toprule
% & & & \multicolumn{3}{c}{Speech} & Audio \\
% \cmidrule(lr){4-6} \cmidrule(lr){7-7}
% Model & VQ Type & Bitrate & ACC(ER)$\uparrow$ & EER(ASV)$\downarrow$ & WER(ASR)$\downarrow$ & mAP(AEC)$\uparrow$ \\
% \midrule
% DAC & RVQ & 6.00 kbps & 72.36\% & 3.94\% & \textbf{4.1\%} & 78.15\% \\
% DAC & RVQ & 4.50 kbps & 71.74\% & 4.46\% & 4.3\% & 74.90\% \\
% % RSimVQ & 70.49\% & 6.23\% & 5.32\% & 75.40\% \\
% BSCodec & 5-band VQ & 3.75 kbps & 67.01\% & 4.01\% & 4.73\% & 82.85\% \\
% BSCodec & 2-band SimVQ & 2.55 kbps & 71.11\% & 4.18\% & 4.62\% & 85.00\% \\
% BSCodec & 3-band SimVQ & 3.83 kbps & \textbf{73.26\%} & \textbf{2.67\%} & 4.4\% & \textbf{87.95\%} \\
% \bottomrule
% \end{tabular}
% \end{table*}

\subsection{Downstream task}
\label{sec:down}

\noindent\textbf{Codec-SUPERB}
We evaluate our model on downstream tasks from the Codec-SUPERB benchmark (SLT Challenge version~\cite{wu2024codec}). We assess performance on three speech tasks: Emotion Recognition (ER, measured by accuracy), Automatic Speaker Verification (ASV, measured by EER), and Automatic Speech Recognition (ASR, measured by WER), as well as Audio Event Classification (AEC, measured by mAP) for the audio domain.

% As shown in Table, our BSCodec with 3-band configuration achieves the best overall performance. The model demonstrates exceptional results on ASV, substantially outperforming both DAC variants. This improvement aligns with the previously observed superior speaker similarity scores, confirming that our band-splitting approach effectively preserves speaker-specific characteristics in low-frequency regions. The model also exhibits outstanding performance on AEC, significantly surpassing both DAC baselines and demonstrating robust preservation of audio event information across diverse acoustic domains. For ER, the 3-band configuration outperforms all baselines, while ASR performance remains competitive across all models. Interestingly, the 5-band configuration shows relatively lower performance on speech tasks, suggesting that overly fine-grained partitioning may fragment speech-relevant information. This reinforces our finding that 3-band partitioning strikes an optimal balance for multi-domain codec design.
As shown in Table~\ref{tab:downstream_results}, our 3-band BSCodec achieves the best performance. It demonstrates exceptional results on ASV, substantially outperforming both DAC variants, aligning with superior speaker similarity scores. The model excels on AEC and ER, significantly surpassing all baselines, while maintaining competitive ASR performance. Notably, the 5-band configuration shows lower performance on speech tasks, suggesting overly fine-grained partitioning fragments speech-relevant information. This confirms 3-band partitioning strikes optimal balance for multi-domain codec design.
\begin{table}[h]
\centering
\footnotesize
\caption{Performance on Codec-SUPERB. The number before \# denotes the number of bands.}
\label{tab:downstream_results}
\begin{tabular}{l@{\hspace{6pt}}c@{\hspace{8pt}}c@{\hspace{8pt}}c@{\hspace{8pt}}c@{\hspace{8pt}}c}
\toprule
& & \multicolumn{3}{c}{Speech} & Audio \\
\cmidrule(lr){3-5} \cmidrule(lr){6-6}
Model & Bitrate & \begin{tabular}{@{}c@{}}ACC$\uparrow$\\(ER)\end{tabular} & \begin{tabular}{@{}c@{}}EER$\downarrow$\\(ASV)\end{tabular} & \begin{tabular}{@{}c@{}}WER$\downarrow$\\(ASR)\end{tabular} & \begin{tabular}{@{}c@{}}mAP$\uparrow$\\(AEC)\end{tabular} \\
\midrule
DAC & 6.00 & 72.36 & 3.94 & \textbf{4.10} & 78.15 \\
DAC & 4.50 & 71.74 & 4.46 & 4.30 & 74.90 \\
BSCodec 5\# & 3.75 & 67.01 & 4.01 & 4.73 & 82.85 \\
BSCodec 2\# & 2.55 & 71.11 & 4.18 & 4.62 & 85.00 \\
BSCodec 3\# & 3.83 & \textbf{73.26} & \textbf{2.67} & 4.40 & \textbf{87.95} \\
\bottomrule
\end{tabular}
\end{table}

\noindent\textbf{ARCH}
We evaluate our model on the ARCH\cite{la2024benchmarking} benchmark, which assesses the semantic richness of codec representations across multiple domains. Following the standard ARCH protocol, we extract and freeze the encoder of each model, append a single linear classification layer, and train for 1000 epochs until full convergence. The speech domain includes the RAVDESS~\cite{livingstone2018ryerson} and Audio-MNIST~\cite{becker2024audiomnist} datasets, the music domain includes the MTT~\cite{law2010evaluation} and MS-DB~\cite{bittner2014medleydb} datasets, and the audio domain includes the ESC50~\cite{piczak2015esc}, US8K~\cite{salamon2014dataset} and VIVAE~\cite{holz2022variably} datasets.

As shown in Table~\ref{tab:classification_results_2}, partitioning into multiple bands with separate encoders substantially enhances the semantic capacity of the encoder. For speech and audio subtasks, finer-grained frequency decomposition enables better capture of domain-specific acoustic characteristics, with the 5-band configuration achieving the best performance. For the music subtask, increasing partitions beyond a moderate level does not yield further improvements, though all band-split configurations significantly outperform the single-encoder baseline. This indicates that while band splitting is universally beneficial for semantic representation learning, the optimal granularity varies across domains, likely reflecting their distinct spectral characteristics and information distribution patterns.

\section{Ablation Study}
% \subsection{Impact of SimVQ}
To eliminate the potential confounding effect of SimVQ itself on the experimental results, we conduct ablation studies using Residual SimVQ and replacing the VQ in band-based quantization with SimVQ. The results are presented in Table~\ref{tab:simvq_ablation}. As shown, neither replacing DAC's RVQ with Residual SimVQ nor substituting VQ with SimVQ in the 5-band BSCodec configuration yields significant performance improvements. In our experiments, SimVQ serves solely as a quantization tool for large codebooks. Moreover, as previously demonstrated, the codebook utilization rate of SimVQ is comparable to that of DAC. Therefore, the design of SimVQ itself is not the decisive factor contributing to the model's superior performance.

\begin{table}[h]
\centering
\small
\caption{Ablation study of VQ method.}
\label{tab:simvq_ablation}
\begin{tabular}{l@{\hspace{2pt}}c@{\hspace{6pt}}c@{\hspace{6pt}}c@{\hspace{6pt}}c}
\hline
\textbf{Model} & \textbf{\#Band} & \begin{tabular}{@{}c@{}}\textbf{Speech}\\\textbf{UTMOS}$\uparrow$\end{tabular} & \begin{tabular}{@{}c@{}}\textbf{Audio}\\\textbf{VISQOL}$\uparrow$\end{tabular} & \begin{tabular}{@{}c@{}}\textbf{Music}\\\textbf{VISQOL}$\uparrow$\end{tabular} \\[2pt]
\hline
\rule{0pt}{3ex}DAC & 1 & 3.201 & 4.055 & 4.171 \\[2pt]
\quad w/ RSimVQ & 1 & 2.655 & 4.185 & 4.267 \\[4pt]
BSCodec & 5 & 2.515 & 4.245 & 4.326 \\[2pt]
\quad w/ SimVQ & 5 & 2.723 & 4.185 & 4.267 \\[2pt]
\hline
\end{tabular}
\end{table}
\vspace{-10pt}
\section{Limitations}
% While our work demonstrates promising results on reconstruction and downstream tasks, some limitations warrant further investigation. Our evaluation primarily focuses on reconstruction quality and certain downstream applications without systematically examining performance on codec-based audio generation.  A more comprehensive evaluation across these applications would offer deeper insights into our model's capabilities and provide clearer directions for future improvements.

The presented BSCodec can further benefit from a more comprehensive evaluation protocol. Current evaluation concentrates on audio reconstruction \ref{sec:musrec} and small-scale understanding-oriented downstream tasks \ref{sec:down}. Further investigation on large-scale codec-based audio generation tasks (e.g., language model-based TTS) can provide a more comprehensive profile of the strength of our BSCodec. 

\section{Conclusion}
We present BSCodec, a band-split neural audio codec for universal audio compression that processes different frequency bands separately through parallel encoder-quantizer-decoder modules, naturally handling the spectral differences between speech, music and sound. Our experiments show that BSCodec achieves better reconstruction quality on music and sound compared to existing codecs while maintaining competitive performance on speech, and downstream task evaluations confirm that the learned representations are effective for audio understanding applications. % This work demonstrates that respecting the natural acoustic structures of different audio types leads to better universal audio compression.

\section*{Acknowledgment}

This work used the Bridges2 at PSC and Delta/DeltaAI NCSA systems through CIS210014 from the ACCESS program, supported by NSF \#2138259, \#2138286, \#2138307, \#2137603, and \#2138296.

\bibliography{custom}

\clearpage

\appendix

\section{Training Convergence Speed Comparison}
\label{app:1}

\begin{figure}[h]
    \centering
    \includegraphics[width=\columnwidth]{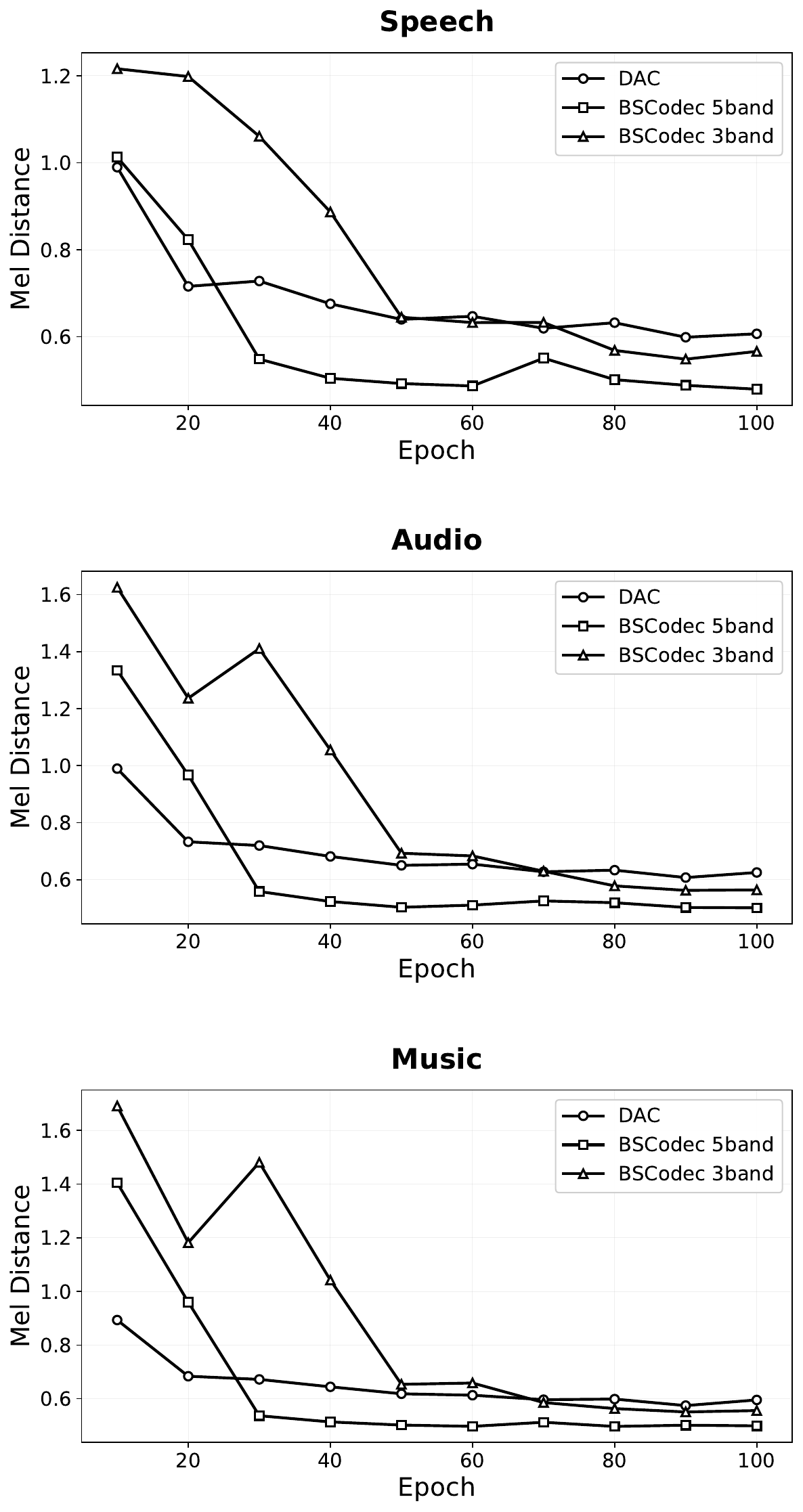}
    \caption{Comparison of the decrease in MEL distance during training for different codecs}
    \label{fig:codec_comparison}
\end{figure}

We compare the convergence speed of different codec architectures by tracking Mel Distance throughout training on the three-domain mixed dataset. Evaluations are conducted every 10 epochs, with each epoch processing 2000 samples.

Figure~\ref{fig:codec_comparison} shows the convergence curves across the three domains. DAC exhibits slower convergence with flatter curves across all domains. Both BSCodec 5-band and 3-band configurations reach stable performance within approximately 40-60 epochs, showing steeper descent in the early training phase.

\section{Codebook Utilization Calculation}
\label{app:2}

For a single-layer codebook with size $K$, the utilization rate is computed by calculating the entropy of the codebook usage distribution and comparing it to the theoretical maximum entropy. Our calculation methodology follows the approach used in DAC~\cite{kumar2023high}. Let $c$ denote the codebook index selected for a given frame, and $p(c)$ be its empirical distribution over the evaluation dataset. The single-layer entropy is:
\begin{equation}
H(c) = -\sum_{j=1}^{K} p(j) \log_2 p(j)
\end{equation}
The single-layer utilization rate is then:
\begin{equation}
U = \frac{H(c)}{\log_2 K}
\end{equation}
where $\log_2 K$ represents the theoretical maximum entropy for a uniform distribution over $K$ entries.

However, this layer-wise approach fails to capture inter-layer dependencies that are critical for understanding the true capacity usage of multi-layer quantization schemes. For a codec with $L$ layers, each having a codebook of size $K$, treating all layers as a single unified codebook yields a theoretical space of size $K^L$. For example, an 8-layer RVQ codec with $K=1024$ per layer corresponds to a joint codebook of size $2^{80}$ (approximately $10^{24}$). Computing the true entropy of such a massive discrete distribution would require statistics over an impractically large audio dataset and is computationally infeasible.

To measure inter-layer correlation while maintaining computational tractability, we adopt a pairwise analysis approach. Specifically, we compute the joint utilization of consecutive layer pairs, treating each pair as a combined codebook of size $K^2$. For layers $i$ and $i+1$, let $c_i$ and $c_{i+1}$ denote the codebook indices selected for a given frame. The joint distribution is:
\begin{equation}
p(c_i, c_{i+1}) = \frac{\text{count}(c_i, c_{i+1})}{N}
\end{equation}
where $N$ is the total number of frames in the evaluation dataset. The joint entropy is:
\begin{equation}
H(c_i, c_{i+1}) = -\sum_{j=1}^{K} \sum_{k=1}^{K} p(j, k) \log_2 p(j, k)
\end{equation}
The pairwise codebook utilization rate is then defined as:
\begin{equation}
U_{i,i+1} = \frac{H(c_i, c_{i+1})}{2\log_2 K}
\end{equation}
where the denominator $2\log_2 K$ represents the theoretical maximum entropy for two independent uniform distributions over $K$ entries. A utilization rate approaching 1.0 indicates that the two layers are nearly statistically independent and fully utilized, while values significantly below 1.0 suggest redundancy or correlation between layers.

For RVQ-based codecs, we compute $U_{i,i+1}$ for all consecutive pairs $(i, i+1)$ where $i \in \{1, 2, \ldots, L-1\}$ to assess the degree of independence across the residual hierarchy.

\section{Speech Reconstruction Metrics}
\label{app:3}
% WB-PESQ STOI UTMOS SPK\_SIM VISQOL 

We provide detailed descriptions of the speech quality metrics used in our evaluation.

\noindent\textbf{Wide-Band Perceptual Evaluation of Speech Quality (WB-PESQ)} is an ITU-T P.862.2 standard metric that predicts subjective listening quality by comparing processed speech to its clean reference. It outputs scores ranging from -0.5 to 4.5, with higher values indicating better perceptual quality.

\noindent\textbf{Short-Time Objective Intelligibility (STOI)} evaluates speech intelligibility by measuring the correlation between short-time temporal envelopes of processed and reference signals. Scores range from 0 to 1, with higher values indicating better intelligibility.

\noindent\textbf{UTMOS (University of Tokyo Mean Opinion Score)} is a deep learning-based metric that predicts the subjective Mean Opinion Score of synthesized speech without human listeners. It provides scores from 1 to 5, where higher scores indicate better perceived quality.

\noindent\textbf{SPK\_SIM (Speaker Similarity)} measures the preservation of speaker characteristics by computing the cosine similarity between speaker embeddings extracted from synthesized and ground-truth utterances using ESPnet-SPK~\cite{jung2024espnet}. Values range from 0 to 1, with higher scores indicating better speaker identity preservation.

\noindent\textbf{VISQOL (Virtual Speech Quality Objective Listener)} is a full-reference metric that predicts perceived audio quality by measuring spectro-temporal similarity between processed and reference signals. Scores range from 1 to 5, with higher values indicating better quality.

\section{Energy Distribution Extraction}
\label{app:energy_extraction}

We extract energy distributions from audio files using the following pipeline:

\noindent\textbf{Preprocessing:} All audio samples are normalized to -23.0 LUFS using pyloudnorm to ensure fair comparison across recordings.

\noindent\textbf{Spectral analysis:} We compute the Short-Time Fourier Transform (STFT) with FFT size $N_{\text{FFT}} = 2048$ and hop length $H = 512$. The energy spectrum is obtained as $E_i(f, t) = |X_i(f, t)|^2$, where $X_i(f, t)$ are the STFT coefficients for file $i$.

\noindent\textbf{Weighted averaging:} For each category (sound, music, speech), we compute the weighted average across all files:
\begin{equation}
    E_{\text{category}}(f) = \frac{\sum_{i=1}^{N} \sum_{t=1}^{T_i} E_i(f, t)}{\sum_{i=1}^{N} T_i}
\end{equation}
where $N$ is the number of files and $T_i$ is the number of frames in file $i$.
The results are plotted on a semi-log scale for 0--8~kHz.

\section{Data Distribution}
\label{app:4}

\begin{table}[htbp]
\centering
\caption{Datasets of Stage 1.}
\label{tab:data_distribution_stage2}
\resizebox{0.48\textwidth}{!}{
\renewcommand{\arraystretch}{1.3}
\begin{tabular}{l c p{6cm}}
\hline
\textbf{Dataset} & \textbf{Duration (hours)} & \textbf{Description} \\
\hline
MTG-Jamendo & 3,764.42 & Open music dataset \\[0.2cm]
MAESTRO & 194.27 & Classical piano performances \newline with aligned MIDI \\[0.2cm]
MUSDB18 & 6.04 & 150 full-length music \newline tracks \\[0.2cm]
\hline
Total & 3,964.73 & \\
\hline
\end{tabular}
}
\end{table}

\begin{table}[h]
\centering
\caption{Datasets of Stage 2.}
\label{tab:data_distribution}
\resizebox{0.48\textwidth}{!}{
\renewcommand{\arraystretch}{1.3}
\begin{tabular}{l c p{6cm}}
\hline
\textbf{Dataset} & \textbf{Duration (hours)} & \textbf{Description} \\
\hline
AudioSet & 742.51 & Large-scale manually \newline annotated audio events \\[0.2cm]
CommonVoice & 14.93 & Multilingual transcribed \newline speech corpus \\[0.2cm]
MTG-Jamendo & 753.82 & Open music dataset \\[0.2cm]
LibriTTS & 528.94 & Multi-speaker English \newline audiobook corpus \\[0.2cm]
MUSDB18 & 6.04 & 150 full-length music \newline tracks \\[0.2cm]
VCTK & 80.80 & 110 English speakers \newline with various accents \\[0.2cm]
\hline
Total & 2,127.04 & \\
\hline
\end{tabular}
}
\end{table}

\end{document}